\documentclass[12pt,preprint]{aastex} 
\usepackage{natbib,graphicx,epstopdf}

\newcommand{\wspitzer}{\emph{Warm-Spitzer}}
\newcommand{\spitzer}{\emph{Spitzer}}

\newcommand{\mearth}{\emph{MEarth}}

\newcommand{\vlt}{\emph{VLT}}

\newcommand{\gjb}{GJ~1214b}

\shorttitle{Broad-band transmission photometry of GJ1214b at 3.6 and 4.5 microns}
\shortauthors{Desert et al.}\def\simgr{\,\hbox{\hbox{$ > $}\kern -0.8em \lower 1.0ex\hbox{$\sim$}}\,}
\def\simle{\,\hbox{\hbox{$ < $}\kern -0.8em \lower 1.0ex\hbox{$\sim$}}\,}

\begin{document}

\title{Observational evidence for a metal rich atmosphere on the super-Earth GJ1214b}

\author{Jean-Michel D\'esert\altaffilmark{1},
Jacob Bean\altaffilmark{1,2},
Eliza Miller-Ricci Kempton\altaffilmark{3},
Zachory K. Berta\altaffilmark{1},
David Charbonneau\altaffilmark{1},
Jonathan Irwin\altaffilmark{1},
Jonathan J. Fortney\altaffilmark{3},
Christopher J. Burke\altaffilmark{1},
Philip Nutzman\altaffilmark{3}
}
\altaffiltext{1}{Harvard-Smithsonian Center for Astrophysics, 60 Garden Street, Cambridge, MA 02138; jdesert@cfa.harvard.edu}
\altaffiltext{2}{Sagan Fellow}
\altaffiltext{3}{Department of Astronomy and Astrophysics, University of California, Santa Cruz, CA 95064, USA}

\begin{abstract}

We report observations of two consecutive transits of the warm super-Earth
exoplanet \gjb\ at 3.6 and 4.5~\micron\ with the Infrared Array Camera
instrument on-board the \spitzer\ Space Telescope. 
The two transit light curves allow for the determination of the transit parameters for this system. 
We find these paremeters to be consistent with the previously determined values and no evidence for transit timing variations.
The main investigation consists of measuring the transit depths in each bandpass to constrain the planet's transmission spectrum. Fixing the system scale and impact parameters, we measure $R_p/R_\star=0.1176^{+0.0008}_{-0.0009}$  and $0.1163^{+0.0010}_{-0.0008}$ at 3.6 and 4.5~\micron, respectively.  
Combining these data with the previously reported \mearth\ Observatory measurements in the red optical 
allows us to rule-out a cloud-free, solar composition (i.e., hydrogen-dominated)
atmosphere at $4.5~\sigma$ confidence. This independently confirms a recent
finding that was based on a measurement of the planet's transmission
spectrum using the \vlt. 
The \spitzer, \mearth, and \vlt\ observations together yield a remarkably flat transmission spectrum over the large wavelength domain spanned by the data.
Consequently, cloud-free atmospheric models require more than
30\% metals (assumed to be in the form of H$_{2}$O) by volume to be
consistent with all the observations.

\end{abstract}


\section{Introduction}\label{intro}

\gjb\ is a transiting super-Earth exoplanet orbiting a nearby M4.5V
star every 1.6 days \citep{charbonneau09} that was discovered by the
\mearth\ Observatory \citep{nutzman08,irwin09}. The first data for this
planet suggested that it has a mass of $6.55\pm0.98~M_{\oplus}$ and a
radius of $2.68\pm0.13~R_{\oplus}$, and subsequent follow-up transit light
curve observations and analysis have confirmed the initial radius estimate
\citep{berta10,carter10,kundurthy10}. Due to degeneracies between possible
interior and atmospheric masses and compositions, a diverse range of
models can reproduce \gjb's mass and radius
\citep{rogers10b,nettelmann10}. This problem had been previously identified for planets in the super-Earth regime \citep{seager07,adams08,rogers10a}. Nevertheless, \gjb\ is larger than
expected for its mass assuming a purely solid composition and mature age; this implies that \gjb\ most probably possesses a significant gas envelope.
An important question then is, what are the characteristics of this
atmosphere?

The transiting nature of \gjb\ fortunately provides the opportunity to study its atmosphere in transmission (e.g. \citealt{charbonneau02}). 
In particular, the overall composition of the planet's atmosphere
(e.g., hydrogen-dominated or metal-rich) can be investigated using
transmission spectroscopy observations through the determination of the
atmospheric scale height \citep{kempton09,kempton10}. 
Because the depth of strong absorption features
in the transmission spectrum is proportional to the atmospheric
scale height, such observations constrain the mean molecular weight of the atmosphere, and thus its bulk composition. Obtaining constraints on the composition of \gjb's atmosphere
can give clues to the planet's origins and interior structure
\citep{rogers10b,nettelmann10}.

In this Letter, we present transit observations of \gjb\ obtained with
\spitzer\ that yield measurements of the planet's transmission spectrum.
Recently, \citet{bean10} also presented a transmission spectrum for the
planet in the red optical from \vlt\ observations. We use our \spitzer\
data along with the \vlt\ and \mearth\ data to determine the composition
of \gjb's atmosphere, and we discuss the results in the context of the
possible interior structure models and origins for the planet.

\section{\wspitzer\ Observations}
\label{sec:spitzer}

\gjb\ was observed for approximately 4.5~h during each of two consecutive
transits with \wspitzer/IRAC \citep{werner04,fazio04} at 3.6 and
4.5~\micron\ (program ID 542). The observations occurred on 26 and 27 April
2010, with the first transit observed at 3.6~\micron. Observations in each visit were gathered in sub-array mode ($32\times32$ pixels) with an exposure time of
1.9~s per frame, which resulted in a 2.0~s cadence. These yielded 109 images
per visit, with 64 frames per image and a total of 6976 frames. Nearly
one-fourth of these exposures were taken during the 53~min planetary transit.

\begin{figure*}[h!]
\begin{center}
 \includegraphics[width=6.5in]{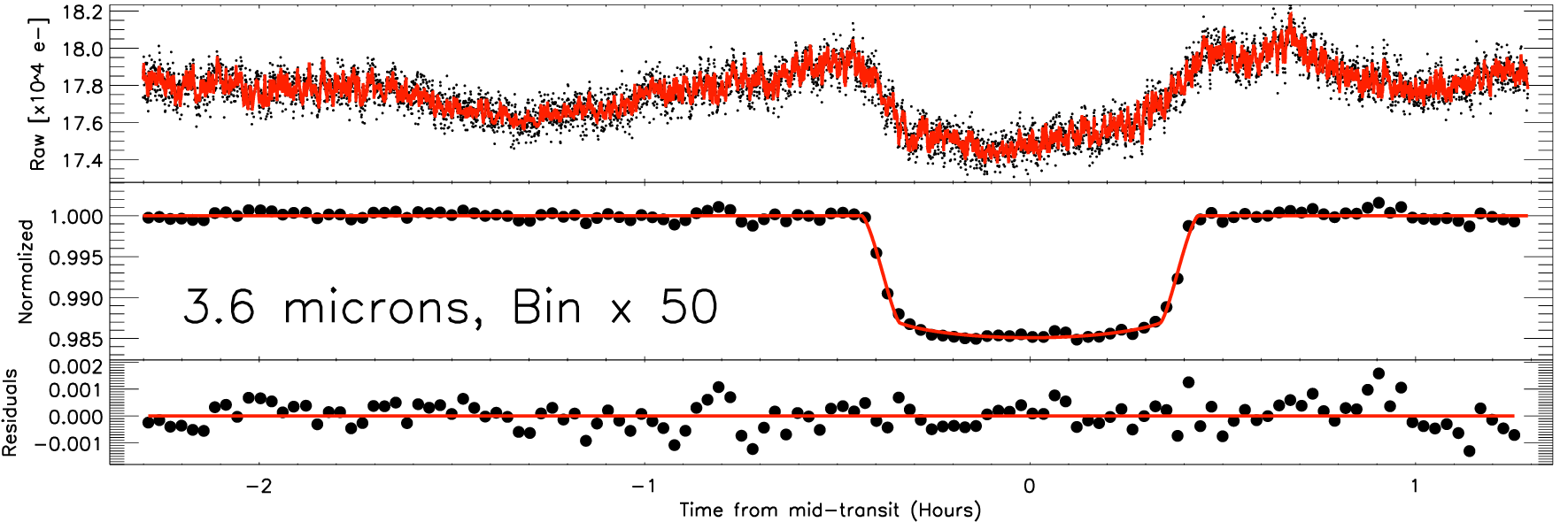}
 \includegraphics[width=6.5in]{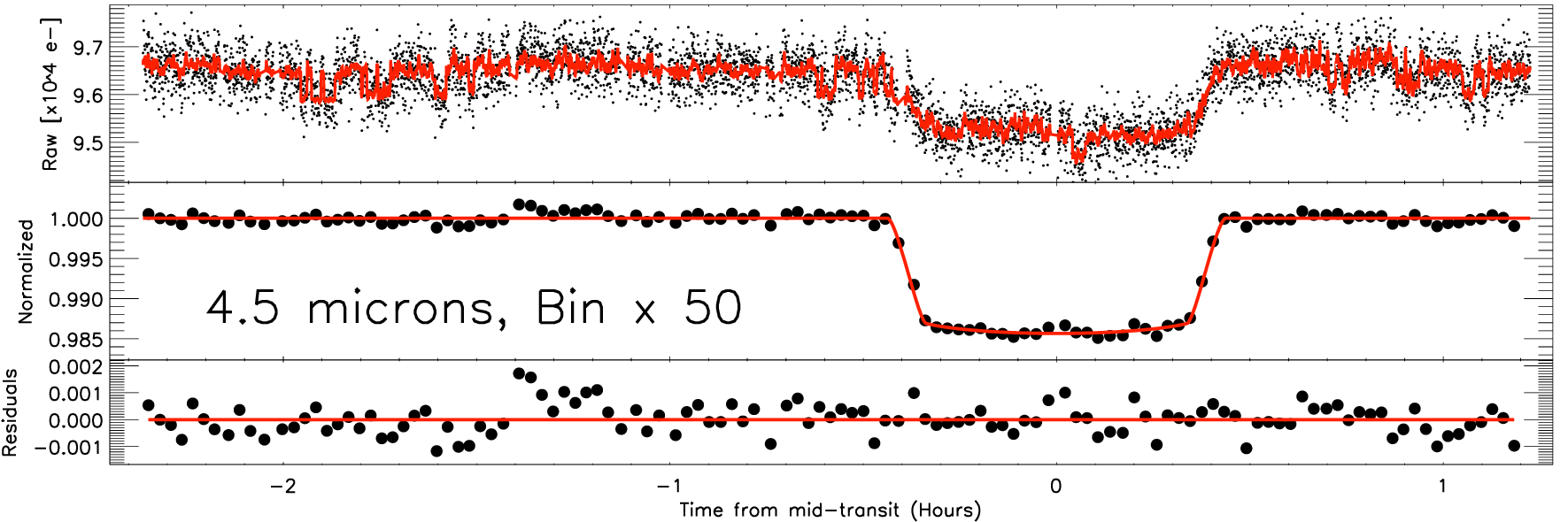}
 \caption{\spitzer\ transit light-curves observed in the two currently available IRAC band-passes at 3.6 (top) and 4.5~\micron\ (bottom). Top panels~: raw transit light-curves. The red solid line corresponds to the best fit models which include the time and position instrumental decorrelations as well as the model for the planetary transit  (see details in Sect.~\ref{sec:spitzer}). Middle panels~: corrected and normalized transit light-curves with their best fits. The data are binned in 100~seconds intervals (50 points). Bottom panels~: residuals.}
   \label{fig:spitzerlightcurves}
\end{center}
\end{figure*}

The method we used to produce photometric time series in each
channel from the Basic Calibrated Data (BCD) is described in \cite{desert10}. It consists of
finding the centroid position of the stellar point spread function (PSF)
and performing aperture photometry. 
The aperture size was optimized to minimize the final residuals in
each channel, although we noticed that the $S/N$ does not vary significantly with the aperture radii. 
The final aperture sizes were set to $5.5$ and
$2.5$~pixels at 3.6 and 4.5~\micron, respectively. The background level
for each frame was determined by fitting a Gaussian to the central region
of a histogram of counts from the full array. We find that the background
varies within the sets of 64 frames, but that it remains globally constant between images and for the whole time-series. 
The $1.9$~s exposures gave a typical signal-to-noise ratio ($S/N$) of $340$ and
$280$ per individual frame  at 3.6 and 4.5~\micron, respectively. 
The noise is $20\%$ and $10\%$ larger than photon noise at 3.6 and  4.5~\micron, respectively.
We used a sliding median filter to select and trim outliers greater than $5~\sigma$, which
correspond to $1\%$ of the data. We also discarded the first
half-hour of observations, which are affected by a significant telescope
jitter before stabilization. The final number of photometric measurements
used are $6194$ and $5998$ for the light curves gathered at 3.6 and
$4.5~\micron$, respectively. The two final raw time series are presented
in Figure~\ref{fig:spitzerlightcurves}.

\section{Determination of the transit parameters}\label{model_transit}
\subsection{Analysis of the \wspitzer\ light curves}

We used a transit light curve model multiplied by instrumental
decorrelation functions to measure the transit parameters and their
uncertainties from the \spitzer\ data as described in \cite{desert10}. We
computed the transit light curves with the IDL transit routine
\texttt{OCCULTQUAD} from \cite{mandel02}, which depends on five
parameters: the planet-to-star radius ratio $R_p / R_\star$, the orbital
semi-major axis to stellar radius ratio (system scale) $a / R_\star$, the
impact parameter $b$, the time of mid transit $T_c$, and a linear limb-darkening coefficient $c_1$.
In the case of late M-dwarfs, such as GJ~1214, the limb-darkening is
uncertain because it strongly depends upon stellar physical parameters
that are not known accurately. Consequently, we opted to determine the
limb-darkening coefficients for our light curve modeling from the data
themselves. We assumed that the limb-darkening is well-approximated by a
linear law \citep{claret98} at infrared wavelengths.

The \spitzer/IRAC photometry is known to be systematically affected by the
so-called \textit{pixel-phase effect} (see e.g., Charbonneau et al. 2005).
This effect is seen as oscillations in the measured fluxes with a period
of approximately 70~min (period of the telescope pointing jitter) and an amplitude of approximately $2\%$
peak-to-peak. 
We found the centroid position of the PSF with \texttt{GCNTRD}, from the IDL Astronomy Library \footnote{{\tt http://idlastro.gsfc.nasa.gov/homepage.html}}.
We decorrelated our signal in each channel using a
linear function of time for the baseline (two parameters) and a quadratic function of the
PSF position (four parameters) to correct the data for each channel as described in
\citet{desert10}.

We performed a simultaneous Levenberg-Marquardt least-squares fit \citep{markwardt09} to the
data to determine the transit and instrumental parameters (11 in total).
The errors on each photometric point were assumed to
be identical, and were set to the $rms$ of the residuals of the initial
best-fit obtained.
To obtain an estimate of the correlated and systematic errors
\citep{pont06} in our measurements, we use the residual permutation bootstrap, or
``Prayer Bead'', method as described in \citet{desert09}. In this method,
the residuals of the initial fit are shifted systematically and
sequentially by one frame, and then added to the transit light curve model
before fitting again. 
We allow asymmetric error bars spanning $34\%$ of the points above and below the median of the distributions to derive the $1~\sigma$ uncertainties for each parameters.

\subsection{Results}\label{results}

We performed two separate analysis of the \spitzer\ data, and the results
are given in Table~\ref{tab:spitzer}. For the first analysis, we fit the
data without any constraints and allowed all the transit parameters to be
unique for each of the two \spitzer\ channels. Our values for all the
parameters are in agreement with the previously reported results
from extensive follow-up observations of \gjb\
\citep{berta10,carter10,kundurthy10,sada10} and the adopted parameters
used to determine the optical transmission spectrum by \citet{bean10}.
Furthermore, the mid-transit times from the \spitzer\ observations are
precisely measured with an accuracy of $\pm 7$~s and do not deviate
significantly from the revised linear ephemeris of \citet{berta10}. This strongly suggests
that the transiting parameters we derive are robust and well constrained
at this point.

The agreement between all the available data sets justifies a second
analysis of the \spitzer\ data, where we fixed the $b$ and $a / R_\star$
values to those used by \citet{bean10} to measure the optical transmission
spectrum while still determining unique $R_p / R_\star$ for each channel and collapsing along the dominant degeneracies between these parameters.
We also reanalyzed the \mearth\ data for five transits presented by
\citet{berta10} with these parameters fixed. This enables us to combine
the \spitzer\ photometry with the \vlt\ spectrum and the \mearth\
photometry to constrain \gjb's transmission spectrum.

Adopting identical $b$ and $a / R_\star$, the \mearth\ data yield
$R_p/R_\star=0.1151^{+0.0007}_{-0.0007}$, and the average radius ratio value in the \vlt\
spectrum is $R_p/R_\star=0.1165$ with a typical uncertainty of 0.0006. For
the \spitzer\ data, we find $R_p/R_\star=0.1176^{+0.0008}_{-0.0009}$ and
$0.1163^{+0.0010}_{-0.0008}$ at 3.6 and 4.5~\micron, respectively (see Table~\ref{tab:spitzer}). As
described by \citet{bean10}, the \vlt\ spectrum is featureless, and we
find that the transit depths derived from the \spitzer, \mearth\ and \vlt\
measurements are also all in agreement within $2~\sigma$ level. 
The linear limb-darkening coefficients we derived for each bandpass are consistent with expected theoretical values \citep{claret00,sing10}.

The period of the telescope jitter leading to the pixel-phase effect has
the same timescale of the transit duration ($1$~hour). Therefore, stronger
degeneracies exist between the transit depths, the linear limb-darkening
coefficients, and the corrections of the intra-pixel sensitivity than is
observed in transit data for planets with longer transit durations (e.g. \citealt{desert09}). This
leads to somewhat larger error bars than had been expected, although
fixing the limb-darkening coefficients only reduces the the error bars on
the transit depths by $20\%$.

Stellar variability can be responsible for time and wavelength dependent
transit depth variations \citep{desert10}.
Since the sets of observations we aim to compare have been secured at
different epochs, it is critical to ensure that GJ~1214's variability does
not affect the measured radius ratios. \citet{berta10} analyzed three
years of photometric monitoring in various bandpasses and found the
stellar rotation period of GJ~1214 to be $52.7\pm5.3$ days, and the
photometric variability in the red optical to be 1\%. This suggests that
the transit depths vary by less than $0.01\%$ and $0.004\%$ (in absolute)
in the \mearth\ and \spitzer\ bandpasses, respectively, which is below the
amplitude errors of the radius ratios. This is also further supported by
multiple transit observations of this object in various bandpasses cited
above. Consequently, the comparison between the transit depth derived from
observations secured at different epoch can be done with no corrections.

\section{Discussion}\label{discussion}

\subsection{The nature of GJ1214b's atmosphere}\label{atmo}

As already mentioned in Sect.~\ref{results}, the wavelength dependent
measurements of $R_p / R_\star$ secured by \spitzer, \mearth, and \vlt\
are all in agreement within the $2~\sigma$ level. In principle, the lack of
features in the wavelength dependent planet radius can be interpreted as
no evidence for the presence of a planetary atmosphere seen in
transmission. This means that either this planet has no atmosphere or its depth variations are below our level of sensitivity. 
However, our current knowledge of the planetary mass and radius rules out the absence of
an atmospheric envelope since it is required by interior structure models of \gjb\ \citep{rogers10b,nettelmann10}. 
Therefore, we assume that \gjb\ possesses a gas envelope in the rest of this paper.

\begin{figure*}[h!]
\begin{center}
\includegraphics[width=7in]{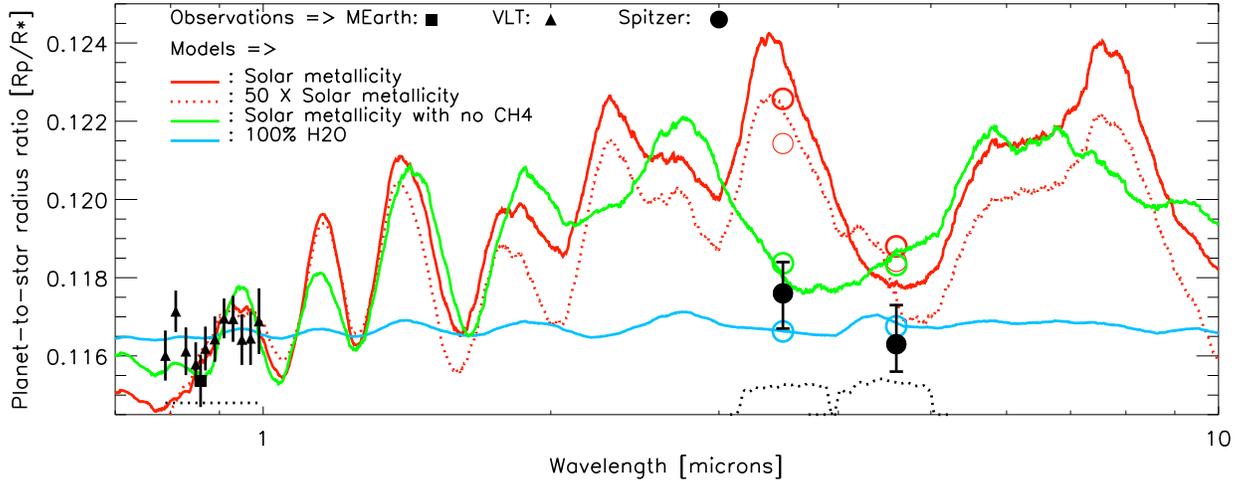}
\caption{Comparison of the  measured wavelength dependent planet-to-star radius ratios to transmission spectroscopic models from \cite{kempton10}. The two radius ratios obtained from the \spitzer\ observations are the black filled circles with their $1~\sigma$ error bars. The results from \mearth\ and \vlt\ observations are shown as black square and triangles, respectively. The continuous lines correspond to the best-scaled and smoothed transmission spectra expected from model of \gjb\ atmosphere's \citep{kempton10}. The red continuous line corresponds to a transmission spectra with solar composition (H/He rich), the red dotted lines is the same spectra but with metallicity enhanced by a factor of 50. The green line is a model with solar composition but with no methane and the blue line is a model with $100\%$ water vapor in the atmosphere. The open circles are the flux weighted integrated models in the \spitzer\ bandpasses. The dotted black lines at the bottom of the plot correspond to the instrumental bandpasses. The hydrogen rich model (red continous line) is ruled out a $7~\sigma$ level by the combined set of observations.}
\label{fig:models}
\end{center}
\end{figure*}

Figure~\ref{fig:models} shows the \spitzer, \mearth, and \vlt\ $R_p /
R_\star$ measurements compared to cloud free transmission spectroscopy models for
\gjb\ with various gas compositions calculated by \citet{kempton10}. 
The atmospheric models solve for the wavelength-dependent deposition of flux, as a function of depth, and make no assumption regarding albedo. 
We assume heat redistribution over the full planet, which results in an equilibrium temperature of 555 K. 
None of these models include cloud formation since abundant condensates are not expected to form at the temperatures expected for the planet's atmosphere.
We consider two main types of atmospheric models. The first type includes
hydrogen-dominated atmospheres with metallicities ranging from solar to 50
times solar, and also a solar composition model with methane artificially
removed. 
The later is considered because methane molecules are susceptible to photolysis since they have large photo-dissociation cross sections in the UV. 
Furthermore, methane does not re-form very quickly through chemical reactions at the temperatures that are expected for the upper atmosphere of \gjb.
The second type of models considered are mixtures of molecular
hydrogen and water (in the form of vapor) ranging from 10 to 100\% water
by volume. All the spectral models were computed at high resolution and were flux weighted integrated over
the bandpasses of the measurements to enable calculation of the $\chi^{2}$
goodness of fit. We
apply a scaling factor to the models to give the best fit to the
measurements.

As a first test, we compare the models to only the \mearth\ and the
\spitzer\ data. We find that these data on their own exclude
hydrogen-dominated atmosphere models with solar metallicity at $4.5~\sigma$
confidence level.
This is due to a strong broad band methane absorption feature at 3.6~\micron\ which disappears in the methane-free model as seen in Figure~\ref{fig:models}.
This is consistent with the findings of \citet{bean10}
using only the \vlt\ spectrum. 
The hydrogen-dominated atmospheres would have a large scale height and lead to large spectral variations that are not observed in the data.
The difference between the radius ratios measured in the two \spitzer\
bandpasses, $\Delta R_p / R_\star=0.0013^{+0.0013}_{-0.0012}$ limits the
range of possible atmospheres because of the methane feature. Of the cloud-free hydrogen-dominated
atmospheres, only those without methane are consistent with the combined \spitzer\ and
\mearth\ data.

The combination of \mearth, \vlt, and \spitzer\ observations provides
stronger constraints on the models, in particular thanks to the spectral
information from the \vlt\ data. The combined data set rules out the model with solar metallicity at $7~\sigma$ confidence level and models with up to 50 times enhanced metallicity at $6.3~\sigma$.
This is due to the sensitivity of the data from the \vlt\ bandpass to spectral
features of water, which has been noted previously by \citet{bean10}.
Including the \vlt\ data also enable us to rule out the methane-free
model at $5~\sigma$ confidence.

The flat transmission spectrum of the combined data
set suggests that the scale height must be small for cloud-free
atmospheres to produce spectral features with an amplitude lower than
$0.02\%$. 
An atmosphere dominated by chemical elements heavier than hydrogen and helium
is therefore required. Although we cannot discriminate between
heavy elements, water in its vapor state is expected to be a dominant
species in various model scenarios \citep{rogers10b,nettelmann10}. The
\mearth, \vlt, and \spitzer\ observations together are consistent within
$3~\sigma$ with atmospheric models containing more than $10\%$ of water by
volume. Models with more than $30\%$ of water are consistent within $1~\sigma$.

An alternative explanation for the featureless transmission spectrum of
\gjb\ indicated by the \mearth, \vlt, and \spitzer\ data could be the
presence of high altitude clouds or hazes. A high altitude cloud deck
would reduce the amplitudes of the expected spectral features because the
stellar light would be transmitted through a smaller atmospheric layer
than in the case of a cloud-free atmosphere. 
The proposed temperature-pressure profiles of the
\citet{kempton10} models do not cross the condensation curves of known equilibrium species that are expected to be abundant \citep{fortney05}. 
Furthermore, clouds and hazes particles have wavelength dependent opacities which seem at odd with the remarkably flat transmission spectrum observed.
However, more theoretical work is needed since, for example, a distribution of particle sizes or photochemical processes, although unconstrained at this point, could modify the transmission signature of this atmosphere.
Hydrocarbon haze formed through photochemical processes and methane depletion by similar processes remain possibilities to explain the broad band observations.
Consequently, we only consider the simple case of a cloud and haze-free atmosphere at chemical equilibrium in the following section.

\subsection{The atmosphere as a diagnostic of the planet's origins and interior}\label{atmo}

\cite{rogers10b} propose three major distinct possible origins for the gas
layer concomitant with interior structure models for \gjb: accretion from
the primordial protoplanetary nebulae, sublimation of ices, or outgassing
of rocky material. Our observations exclude the primordial atmosphere scenario (with possible exceptions such as a hazy methane-free atmosphere) because
this would yield a hydrogen-dominated composition. Atmospheres arising from
sublimation or outgassing are consistent with the data.

If sublimation of ices is the atmospheric origin scenario, it would
suggest that the planet has a high concentration of water in its interior.
This implies that the planet formed beyond the snow line of its host
star's protoplanetary disk, and that it migrated inward to its current
orbital position. After the planet migrated closer to its star, the ice
forming the bulk sublimated and led to a metal-dominated atmosphere as it
is currently observed. There are two possible scenarios to explain the
hydrogen-poor envelope in this case. Either the planet never attained a
mass sufficient to accrete or retain large amounts of primordial gas
(H-rich) from the protoplanetary disk \citep{rafikov06}, or the mass
fractions of light elements have been diminished through atmospheric
escape. \cite{rogers10b} point out that an atmosphere with a small scale
height, such as suggested by our observations, is more robust against
atmospheric escape than a hydrogen-rich envelop, and that the cumulative
loss of mass from a water-dominated atmosphere would not significantly
affect the overall character of the planet. 
Interestingly, \citet{nettelmann10} show that a water-rich planet that completely lacks
hydrogen in the atmosphere requires an implausibly large water-to-rock
ratio. Instead, H/He/H$_{2}$O envelopes with water mass fractions between
approximately 50 and 85\% are favored. This picture is in agreement with the \mearth, \vlt, and \spitzer\ observations, which
suggest a water mass fraction in the atmosphere of more than $50\%$.

The alternative scenario for the origin of \gjb's atmosphere would require a giant
terrestrial-like planet that experienced dissolution of icy planetesimals into the envelope (e.g. \citealt{pollack96}), or outgassing during formation and evolution or a period of tectonic activity \citep{schaefer07,elkins08,schaefer09,kite09}. \cite{rogers10b} discussed
this scenario in the context of an atmosphere composed of purely hydrogen
on the grounds that heavy elements like water or carbon dioxide can not be
out-gassed in sufficient quantity to form an atmosphere that can account
for \gjb's large size on their own. However, they did not consider the
possibility of a hybrid atmosphere with significant amounts of both
hydrogen and heavy elements. While heavy elements alone can not be
out-gassed in sufficient quantities, \citet{elkins08} have shown that
significant amounts of water can be out-gassed from chondritic
planetesimals. Given this, and that \citet{nettelmann10} have indicated a
mixed H/He/H$_{2}$O envelope can successfully be used to model the planet,
a hybrid composition atmosphere from outgassing seems to be a plausible
scenario. 


\begin{deluxetable}{lcccccc}
\tabletypesize{\scriptsize}
\tablecaption{\gjb\ parameters and uncertainties from \spitzer\ observations.}
\tablewidth{0pt}
\tablehead{\colhead{Wavelength} & \colhead{$R_p/R_\star$} & \colhead{$b$} & \colhead{$a/R_\star$} & \colhead{$c_{1}$} & \colhead{Tc (BJD$_{BTD}$) \tablenotemark{a}} & \colhead{O-C (Sec.) \tablenotemark{b}}}

\startdata
3.6~\micron\ &  $0.1176^{+0.0010}_{-0.0010}$ & $0.10^{+0.17}_{-0.10}$ & $15.46^{+0.13}_{-0.47}$ & $0.218^{+0.054}_{-0.050}$ & $55312.633877^{+7.6e-05}_{-8.5e-05}$ & $-13^{+12}_{-12}$ \\

4.5~\micron\ &  $0.1171^{+0.0011}_{-0.0015}$ & $0.25^{+0.11}_{-0.16}$ & $15.25^{+0.39}_{-0.58}$ & $0.146^{+0.088}_{-0.051}$ & $55314.214231^{+8.1e-05}_{-9.8e-05}$ & $-11^{+12}_{-12}$ \\

     \\
\hline              
     \\

3.6~\micron\ &  $0.1176^{+0.0008}_{-0.0009}$ & fixed to 0.27729 & fixed to 14.9749  & $0.239^{+0.044}_{-0.061}$ & $2455312.633085^{+8.0e-05}_{-7.6e-05}$  & $-10^{+12}_{-12}$ \\

4.5~\micron\ &  $0.1163^{+0.0010}_{-0.0008}$ &  fixed to 0.27729 &  fixed to 14.9749 &  $0.305^{+0.108}_{-0.067}$ & $2455314.213440^{+8.2e-05}_{-9.3e-05}$ & $-9^{+12}_{-12}$ \\
\enddata
\tablenotetext{a}{Times are given as Barycentric Julian Dates in the Barycentric
Dynamical Time system \citep{eastman10}.}
\tablenotetext{b}{Observed minus Calculated (O-C) using the ephemeris from \citep{berta10} with Period (days)=$1.58040490\pm 0.00000033$ and T$_{0}$ (BJD$_{BDT}$)=$2454966.525042\pm 0.000065$.}
\label{tab:spitzer}
\end{deluxetable}


\clearpage

\acknowledgments
We thank Bryce Croll, Heather Knutson, Dimitar Sasselov, Nadine Nettelmann and Leslie Rogers for useful discussions.
This work is based on observations made with the Spitzer Space Telescope which is operated by the Jet Propulsion Laboratory, California Institute of Technology under a contract with NASA. 
We thank the Sagan Fellowship Program, which provides support for JB and EK.

\end{document}